\documentstyle[eqsecnum,prc,aps]{revtex}
\begin{document}
\draft

\title{Maximally incompressible neutron star matter}
\author{Timothy S. Olson\cite{byline}}
\address{Salish Kootenai College, P.O. Box 117, Pablo, Montana 59855}
\date{4 October 2000}
\maketitle

\begin{abstract}
Relativistic kinetic theory, based on the Grad method of moments as developed by Israel and Stewart, is used to model viscous and thermal dissipation in neutron star matter and determine an upper limit on the maximum mass of neutron stars. In the context of kinetic theory, the equation of state must satisfy a set of constraints in order for the equilibrium states of the fluid to be thermodynamically stable and for perturbations from equilibrium to propagate causally via hyperbolic equations. Application of these constraints to neutron star matter restricts the stiffness of the most incompressible equation of state compatible with causality to be softer than the maximally incompressible equation of state that results from requiring the adiabatic sound speed to not exceed the speed of light. Using three equations of state based on experimental nucleon-nucleon scattering data and properties of light nuclei up to twice normal nuclear energy density, and the kinetic theory maximally incompressible equation of state at higher density, an upper limit on the maximum mass of neutron stars averaging 2.64 solar masses is derived.
\end{abstract}

\twocolumn

\section{introduction}
The problem of the maximum possible mass of neutron stars has a long history. Baade and Zwicky in 1934 proposed a star composed of densely packed neutrons as the final state resulting from the supernova process \cite{Baade34}. Oppenheimer and Volkoff in 1939 demonstrated a star composed of noninteracting neutrons is supported against gravitational collapse to a black hole by the Fermi degeneracy pressure only for stellar masses up to 0.72 solar masses ($M_\odot$) \cite{Oppenheimer39}. The discovery of the first pulsar in 1967 and the realization pulsars are highly magnetized rotating neutron stars triggered substantial theoretical work towards understanding the structure of neutron stars. Pulsar masses can be measured from observations of Doppler shifts of periodic emissions from neutron stars in binary systems. Some 20 radio pulsar masses have been accurately determined so far, with most masses clustering around 1.4 $M_\odot$ \cite{Thorsett99}. This is well beyond the value of 0.72 $M_\odot$ computed for noninteracting neutrons, so the short-range repulsion of the strong nuclear force must make a substantial contribution to the pressure supporting neutron stars against gravitational collapse. There is also evidence for neutron stars with substantially larger masses than the radio pulsars due to accretion of matter from a binary companion. The mass of the x-ray pulsar in the Vela X-1 binary has been estimated to be 1.9 $M_\odot$ \cite{vanKerkwijk00}, and the x-ray burster Cygnus X-2 has been determined to be 1.8 $M_\odot$ \cite{Orosz00}. Quasiperiodic oscillations in the x-ray emissions from neutron stars accreting matter from low mass companions have been argued to imply a mass of up to 2.3 $M_\odot$ for the neutron stars in these systems \cite{Zhang97,Miller98}.

An accurate theoretical determination of the maximum possible mass of neutron stars is of practical interest for identifying as a black hole any compact object with a larger mass. The maximum mass value remains uncertain because the equation of state of neutron star matter is not well understood at the high density values found in neutron star interiors. Modern methods develop the equation of state by fitting experimental nucleon-nucleon scattering data and properties of light nuclei to two and three-body interaction potentials \cite{Wiringa88,Akmal98}. Confidence in the results of these methods is high near normal nuclear density (energy density = 152 MeV/fm$^3$ = 2.7 X 10$^{14}$ g/cm$^3$, baryon density = 0.16 fm$^{-3}$) because, at low density, experimental verification is possible with existing facilities. The validity of extrapolating the experimentally verified low density equation of state to high density, or as an alternative theoretically modeling the physical processes thought to occur at high density,  is uncertain because of the lack of high density laboratory data. The sources of the uncertainties in the high density equation of state are incomplete knowledge of the three-nucleon interactions, the contributions of mesonic and other baryonic besides nucleonic degrees of freedom, and a possible quark deconfinement transition \cite{Akmal98}. 

In light of these uncertainties in the exact form of the high density equation of state, several authors have derived an upper limit on the maximum neutron star mass using only general restrictions on the equation of state. Oppenheimer and Volkoff suggested \cite{Oppenheimer39}, and Rhoades and Ruffini rigorously proved \cite{Rhoades74}, the maximum mass of a stable neutron star results when the stiffest equation of state compatible with thermodynamic stability and causality is used. Rhoades and Ruffini derived an upper limit on the maximum neutron star mass of 3.2 $M_\odot$ by using the experimentally verified Harrison-Wheeler equation of state \cite{Harrison65} up to 1.7 times normal nuclear energy density, and the stiffest equation of state compatible with causality at high density. They defined a casual equation of state to be one having an adiabatic sound speed less than the speed of light, so they took the most incompressible equation of state to have  the sound speed equal to light speed. Kalogera and Baym \cite{Kalogera96} updated the result of Rhoades and Ruffini, using an equation of state developed by Wiringa, Fiks, and Fabrocini \cite{Wiringa88} up to twice normal nuclear energy density. They derived an upper mass limit of 2.9 $M_\odot$ using the same maximally incompressible equation of state as Rhoades and Ruffini above twice normal nuclear energy density.

The purpose of this paper is to apply insights from relativistic kinetic theory to the problem of the maximum possible mass of neutron stars. Kinetic theory used to model viscosity and heat conduction shows requiring the sound speed to be less than the speed of light is a necessary but not sufficient condition for causality. The equation of state must satisfy a set of constraining conditions for the fluid to have stable equilibrium states with perturbations from equilbrium that propagate causally via a system of hyperbolic equations \cite{Hiscock83,Olson90}. If any one of these constraints is violated, there will be at least one fluid mode with a superluminal propagation speed. In particular, it is possible for the fluid to be acausal in a transverse mode or a different longitudinal mode even though the sound speed is below light speed. These constraints generally fix the stiffest possible equation of state to have an adiabatic sound speed significantly less than the speed of light. The main result of the present paper is application of these constraints to neutron star matter at high density softens the maximally incompressible equation of state and reduces the maximum neutron star mass significantly below the value found using a high density equation of state having the adiabatic sound speed equal to the speed of light. In the next section the thermodynamic constraint conditions from relativistic kinetic theory are reviewed. Then the maximally incompressible equation of state of neutron star matter is developed from these constraint conditions. Finally, neutron star models are constructed using this stiffest possible equation of state at high density, and the resulting maximum neutron star mass is determined. Gravitational units having $G=c=1$ are used.

\section{relativistic kinetic theory constraints on the equation of state}
The influences of dissipation in the form of bulk and shear viscosity and thermal conductivity in neutron star matter are modeled here using relativistic kinetic theory based on the Grad method of moments \cite{Grad49} as developed by Israel and Stewart \cite{Israel76,Stewart77,Israel79a,Israel79b}. Reference \cite{Olson89} contains a concise overview of the Israel-Stewart theory. The simpler relativistic fluid theories of Eckart \cite{Eckart40} and Landau-Lifshitz \cite{Landau59} are not appropriate for this problem because they have been demonstrated to exhibit pathological properties. These theories do not possess stable equilibrium states regardless of the equation of state used \cite{Hiscock85}, and there are fluid modes having propagation speeds greater than the speed of light \cite{Hiscock87}.

The Israel-Stewart theory has been shown to possess stable equilibrium states, and linear perturbations from equilibrium propagate casually via a hyperbolic system of equations and form a well-posed initial value problem \cite{Hiscock83,Olson90,Hiscock88}. The thermodynamic variables describing the fluid must satisfy a set of conditions in order for the fluid to be thermodynamically stable, causal, and hyperbolic:
\begin{equation} \label{Omega1}
\frac{1}{\rho + p} \left( \frac{\partial \rho}{\partial p} \right)_s \geq 0 \quad ,
\end{equation}
\begin{equation}
\frac{1}{\rho + p} \left( \frac{\partial \rho}{\partial s} \right)_p \left( \frac{\partial p}{\partial s} \right)_\Theta \geq 0 \quad ,
\end{equation}
\begin{equation}
\left( \rho + p \right) \left[ 1 - \left( \frac{\partial p}{\partial \rho} \right)_s \right] - \frac{1}{\beta_0} - \frac{2}{3\beta_2} - \frac{K^2}{\Omega} \geq 0 \quad ,
\end{equation}
\begin{equation}
\rho + p - \frac{\beta_1 + 2\beta_2 + 2\alpha_1}{2\beta_1\beta_2 - \alpha_1^2} \geq 0 \quad ,
\end{equation}
\begin{equation}
\beta_0 \geq 0 \quad ,
\end{equation}
\begin{equation}
\Omega \geq 0 \quad ,
\end{equation}
\begin{equation}
\beta_1 - \frac{\alpha_1^2}{2\beta_2} \geq 0 \quad ,
\end{equation}
\begin{equation} \label{Omega8}
\beta_2 \geq 0 \quad ,
\end{equation}
where
\begin{equation}
K = 1 + \frac{\alpha_0}{\beta_0} + \frac{2\alpha_1}{3\beta_2} - \frac{n}{T} \left( \frac{\partial T}{\partial n} \right)_s \quad ,
\end{equation}
\begin{equation} \label{Omega6}
\Omega = \beta_1 - \frac{\alpha_0^2}{\beta_0} - \frac{2\alpha_1^2}{3\beta_2} - \frac{1}{nT^2}\left( \frac{\partial T}{\partial s} \right)_n \quad .
\end{equation}
Satisfaction of these conditions is necessary and sufficient for stability, causality, and hyperbolicity of linear perturbations. In Eqs. (\ref{Omega1})-(\ref{Omega6}), $\rho, p, n, T$, and $s$ are, respectively, the energy density (including mass-energy), pressure, particle density, temperature, and entropy per particle, each as measured in a comoving frame. The thermodynamic potential $\Theta$ is the relativistic chemical potential divided by the temperature:
\begin{equation}
\Theta = \frac{\rho + p}{nT} - s \quad .
\end{equation}
$\beta_1, \beta_2$, and $\beta_3$ are effectively relaxation times for, respectively, bulk viscous, thermal, and shear viscous dissipation. $\alpha_0$ is the strength of the bulk stress-heat flow vector coupling, and $\alpha_1$ is the shear stress-heat flow vector coupling strength (see Ref. \cite{Olson89} for details). A complete fluid description requires specification of an equation of state and functional forms for the bulk and shear viscosities, thermal conductivity, relaxation times, and bulk-thermal and shear-thermal couplings. Whether or not a fluid state is thermodynamically stable, causal, and hyperbolic is determined solely by the equation of state, the relaxation times, and the viscous stress-heat flow vector coupling strengths.

For neutron star matter it is convenient to choose the particle (baryon) density $n$ and the temperature $T$ as the independent thermodynamic variables. The energy density and pressure can then be written as
\begin{equation} \label{nuclearEnergyDensity}
\rho(n,T) = n\left[ m + E_0(n) + I(n,T) \right]
\end{equation}
and
\begin{equation} \label{nuclearPressure}
p(n,T)=n^2\left[ \frac{dE_0}{dn} + \left( \frac{\partial I}{\partial n}\right)_s \right] \quad .
\end{equation}
In neutron stars the thermal energy per baryon $I$ is small compared to the baryon rest mass energy $m$ and the ground state energy per baryon $E_0$.

The functional forms for the relaxation times and the viscous-thermal coupling strengths are currently unknown for fully interacting nuclear matter. Their forms have been calculated for noninteracting degenerate Fermi gases \cite{Israel79b,Olson89}:
\begin{equation} \label{alpha0}
\alpha_0 = \frac{9\left( 1 + \nu \right) \left( \nu^2 + 2\nu + 2 \right) \left( \beta \nu \right)^2}{A_0 \pi^2 \nu^2 \left( \nu^2 + 2\nu \right)^{3/2}} + O\left( \beta \nu \right)^0 \quad ,
\end{equation}
\begin{equation}
\alpha_1 = -\frac{6\left( 1 + \nu \right)}{A_0 \left( \nu^2 + 2\nu \right)^{5/2}} + O\left( \beta \nu \right)^{-2} \quad ,
\end{equation}
\begin{equation}
\beta_0 = \frac{405\left( 1 + \nu \right)^5 \left( \beta \nu \right)^4}{16 A_0 \pi^4 \nu^4 \left( \nu^2 + 2\nu \right)^{1/2}} + O\left( \beta \nu \right)^2 \quad ,
\end{equation}
\begin{equation}
\beta_1 = \frac{9\left( 1 + \nu \right) \left( \beta \nu \right)^2}{A_0 \pi^2 \nu^2 \left( \nu^2 + 2\nu \right)^{3/2}} + O\left( \beta \nu \right)^0 \quad ,
\end{equation}
\begin{equation} \label{beta2}
\beta_2 = \frac{15\left( 1 + \nu \right)}{2 A_0 \left( \nu^2 + 2\nu \right)^{5/2}} + O\left( \beta \nu \right)^{-2} \quad ,
\end{equation}
where
\begin{equation}
A_0 = \frac{m^4 g}{2\pi^2 \hbar^3}
\end{equation}
and $g$ is the spin weight. Equations (\ref{alpha0})-(\ref{beta2}) are valid for any inverse dimensionless temperature ($\beta = m/kT$) provided the degenerate limit $\beta \nu >> 1$ is satisfied. The dimensionless potential $\nu$ is the nonrelativistic chemical potential per particle divided by the baryon rest energy:
\begin{equation}
\nu = \frac{\rho + p}{nm} - \frac{s}{k \beta} - 1 \quad .
\end{equation}
The shear stress relaxation time Eq. (\ref{beta2}) has recently been used to show the coupling predicted by kinetic theory between vorticity and shear viscous stress may cause colder neutron stars undergoing accertion spin-up to remain stable against $r$-mode instabilities at higher spin rates \cite{Rezania00}. Here the noninteracting degenerate Fermi gas forms will also be used to approximate the relaxation times and viscous-thermal couplings for neutron star matter. The thermal energy per baryon $I$ is poorly understood for neutron star matter, and the noninteracting degenerate Fermi gas form (which is proportional to $1/\beta^2$) will be used.

When Eqs. (\ref{alpha0})-(\ref{beta2}) are substituted into Eqs. (\ref{Omega1})-(\ref{Omega8}) and the low temperature limit $\beta \rightarrow \infty$ is taken, the result is
\begin{equation} \label{densityLimit}
\rho \geq 0 \quad ,
\end{equation}
\begin{equation}
p \geq 0 \quad ,
\end{equation}
\begin{equation} \label{soundSpeedLowerLimit}
\left( \rho + p \right) v_s^2 \geq 0 \quad ,
\end{equation}
\begin{equation} \label{soundSpeedUpperLimit}
v_s^2 \leq \frac{\rho - \frac{1}{3}p}{\rho + p} \quad ,
\end{equation}
\begin{equation}
p \leq 3\rho \quad . \label{pressureUpperLimit}
\end{equation}
If any of Eqs. (\ref{densityLimit})-(\ref{pressureUpperLimit}) are violated, kinetic theory predicts the corresponding equilibrium state will be unstable and the fluid will possess at least one mode with a propagation speed exceeeding light speed. Both the energy density and pressure must be nonnegative for stability. Further, the adiabatic sound speed $v_s \left[v_s^2 = \left( \partial p/\partial \rho \right)_s\right]$ must be real so that pressure does not decrease with increasing density: if Eq. (\ref{soundSpeedLowerLimit}) is violated, small elements of matter would be predicted to spontaneously collapse. The equation of state of neutron star matter was assumed to satisfy Eqs. (\ref{densityLimit})-(\ref{soundSpeedLowerLimit}), along with $v_s^2 \leq 1$, by Rhoades and Ruffini \cite{Rhoades74} and Kalogera and Baym \cite{Kalogera96} in deriving the maximum neutron star mass limit. Here the adiabatic sound speed will be required to satisfy Eq. (\ref{soundSpeedUpperLimit}) rather than $v_s^2 \leq 1$. Equation (\ref{soundSpeedUpperLimit}) implies the fluid state is acausal, even though the adiabatic sound speed is less than the speed of light, when
\begin{equation}
\frac{\rho - \frac{1}{3}p}{\rho + p} < v_s^2 < 1 \quad .
\end{equation}
Equation (\ref{pressureUpperLimit}) does not affect the value of the maximum neutron star mass because it is not violated until the central density is well above the value yielding the maximum mass limit.

For neutron stars the thermal energy can be neglected and Eqs. (\ref{densityLimit})-(\ref{pressureUpperLimit}) expressed solely in terms of the ground state energy per baryon and the rest energy:
\begin{equation}
m + E_0 \geq 0 \quad ,
\end{equation}
\begin{equation}
\frac{dE_0}{dn} \geq 0 \quad ,
\end{equation}
\begin{equation}
n\frac{d^2E_0}{dn^2} - 2\frac{dE_0}{dn} \geq 0 \quad .
\end{equation}
\begin{equation}
m + E_0 - n^2\frac{d^2E_0}{dn^2} - \frac{7n}{3}\frac{dE_0}{dn} \geq 0 \quad ,
\end{equation}
\begin{equation}
3m + 3E_0 - n\frac{dE_0}{dn} \geq 0 \quad ,
\end{equation}
The zero temperature adiabatic sound speed can be written as
\begin{equation} \label{soundSpeed}
v_s^2 = \frac{n^2\frac{d^2E_0}{dn^2} - 2n\frac{dE_0}{dn}}{m + E_0 + n\frac{dE_0}{dn}} \quad .
\end{equation}

\section{maximally incompressible equation of state}
In a fully relativistic theory of neutron star matter fluid perturbations would propagate causally. One mode for propagation of fluid perturbations is via adiabatic sound waves, and many of the equations of state which are based on fitting experimental nucleon scattering data predict sound waves will travel faster than the speed of light at high density. For example, the beta-stable A18 + $\delta v$ + UIX$^*$ equation of state of Akmal, Panharipande, and Ravenhall \cite{Akmal98} has an adiabatic sound speed that is greater than the speed of light above $n = 0.86$ fm$^{-3}$. At zero temperature the pressure is a function of only the energy density, and so $v_s^2 = dp/d\rho$. The pressure must not increase with increasing energy density faster than
\begin{equation} \label{oldMaxEOS}
p(\rho) = \rho + \text{const}
\end{equation}
if the adiabatic sound speed is to not exceed light speed. Using the expressions for the energy density and pressure written as functions of the ground state energy per baryon [Eqs. (\ref{nuclearEnergyDensity}) and (\ref{nuclearPressure})], Eq. (\ref{oldMaxEOS}) at zero temperature can be written as a first-order ordinary differential equation for $E_0$:
\begin{equation} \label{oldEnergyDiffEq}
n^2 \frac{dE_0}{dn} = n\left( m + E_0 \right) + \text{const} \quad .
\end{equation}
The general solution of Eq. (\ref{oldEnergyDiffEq}) is
\begin{equation} \label{oldMaxIncompressible}
E_0(n) = C_1 \left( \frac{n}{\tilde{n}} \right) + C_2 \left( \frac{n}{\tilde{n}} \right)^{-1} - m \quad .
\end{equation}
Equation (\ref{oldMaxIncompressible}) can be regarded as the maximally incompressible equation of state consistent with a subluminal sound speed.

Relativistic kinetic theory has shown an adiabatic sound speed less than the speed of light is a necessary but not sufficient condition for the causal propagation of fluid perturbations. The equation of state must satisfy Eqs. (\ref{densityLimit})-(\ref{pressureUpperLimit}) for all densities relevant for neutron star models for the equation of state to be causal. In this paper the maximally incompressible equation of state is taken as the form that satisfies Eq. (\ref{soundSpeedUpperLimit}) with an equality:
\begin{equation} \label{maxSoundSpeed}
v_s^2 = \frac{\rho - \frac{1}{3}p}{\rho + p} \quad .
\end{equation}
Equations (\ref{nuclearEnergyDensity}), (\ref{nuclearPressure}), and (\ref{soundSpeed}) can be used to write Eq. (\ref{maxSoundSpeed}) as a second-order ordinary differential equation for $E_0$:

\begin{equation} \label{newEnergyDiffEq}
n^2 \frac{d^2 E_0}{dn^2} + \frac{7}{3}n\frac{dE_0}{dn} - E_0 = m \quad .
\end{equation}
The corresponding differential equation for the total energy per baryon $E = m_0 + E_0$ is homogeneous:
\begin{equation}
n^2 \frac{d^2 E}{dn^2} + \frac{7}{3}n\frac{dE}{dn} - E = 0 \quad .
\end{equation}
This is a form of the Cauchy linear equation. The substitution $n = e^z$ results in
\begin{equation}
\frac{d^2 E}{dz^2} + \frac{4}{3}\frac{dE}{dz} - E = 0 \quad ,
\end{equation}
which has constant coefficients and can be solved for $E(z)$ with standard methods. After the substitution $z = ln(n)$ the general solution to Eq. (\ref{newEnergyDiffEq}) results in
\begin{eqnarray} 
E_0(n) &&= D_1 \left( \frac{n}{\tilde{n}} \right)^{\left( \sqrt{13}-2 \right) /3} + D_2 \left( \frac{n}{\tilde{n}} \right)^{-\left( \sqrt{13}+2 \right) /3} - m \nonumber \\
&&\approx D_1 \left( \frac{n}{\tilde{n}} \right)^{0.535} + D_2 \left( \frac{n}{\tilde{n}} \right)^{-1.869} - m \label{newMaxIncompressible}
\quad .
\end{eqnarray}
Equation (\ref{newMaxIncompressible}) can be regarded as the maximally incompressible equation of state compatible with the causality condition Eq. (\ref{soundSpeedUpperLimit}). Equation (\ref{newMaxIncompressible}) predicts the ground state energy increases more slowly with increasing baryon density than for Eq. (\ref{oldMaxIncompressible}), so the pressure, which is proportional to the rate of change of the ground state energy, will increase more slowly. Hence Eq. (\ref{newMaxIncompressible}) is a softer maximally incompressible equation of state than Eq. (\ref{oldMaxIncompressible}) and will less readily support the star against gravitational collapse, hence yielding a lower maximum neutron star mass limit than Eq. (\ref{oldMaxIncompressible}).

\section{maximum mass of neutron stars}
In this section the kinetic theory maximally incompressible equation of state Eq. (\ref{newMaxIncompressible}) is used to determine a new least upper bound on the maximum neutron star mass stable against gravitational collapse to a black hole. An experimentally verified equation of state is used at low density, and the maximally incompressible equation of state Eq. (\ref{newMaxIncompressible}) is used at high density and compared to using the sound speed equals light speed maximally incompressible equation of state Eq. (\ref{oldMaxIncompressible}). The low density equation of state consists of the BPS equation of state \cite{Baym71} below normal nuclear density, as in Ref. \cite{Kalogera96}, and one of the recently developed equations of state which utilize fits of nucleon scattering data and light nuclei properties to cover the density range above normal nuclear density. Three of these experimentally based equations of state will be used here as representative examples: the beta-stable AV14 + UVII and the beta-stable UV14 + UVII models of Ref. \cite{Wiringa88} (hereafter WFF-A and WFF-U), and the beta-stable A18 + $\delta v$ + UIX* model of Ref. \cite{Akmal98} (hereafter APR). The maximum mass value is insensitive to the particular equation of state applied below normal nuclear density because little of the stellar mass is in the subnuclear crust when the star is near the mass limit. The low and high density equations of state are matched at a baryon density $\tilde{n}$ by choosing the integration constants $C_i$ and $D_i$ in Eqs. (\ref{oldMaxIncompressible}) and (\ref{newMaxIncompressible}) so that the ground state energy and pressure values of the low and high density equations of state have the same value at the match point. For example, using the WFF-A equation of state at low density and choosing the match point to be twice normal nuclear energy density (304 MeV/fm$^3$, 0.317 baryons/fm$^3$), the ground state energy and pressure at the match point are $E_0(\tilde{n}) = 21.9$ MeV and $p(\tilde{n}) = 8.67$ MeV/fm$^3$. The corresponding values of the $C_i$ and $D_i$ in the maximally incompressible equations of state are
\begin{equation}
C_1 = \frac{\rho(\tilde{n}) + p(\tilde{n})}{2\tilde{n}} \approx 494 \text{ MeV} \quad ,
\end{equation}
\begin{equation}
C_2 = \frac{\rho(\tilde{n}) - p(\tilde{n})}{2\tilde{n}} \approx 467 \text{ MeV} \quad ,
\end{equation}
\begin{equation}
D_1 = \frac{\left( 2 + \sqrt{13} \right) \rho(\tilde{n}) + 3p(\tilde{n})}{2\sqrt{13} \tilde{n}} \approx 759 \text{ MeV} \quad ,
\end{equation}
\begin{equation}
D_2 = - \frac{\left( 2 - \sqrt{13} \right) \rho(\tilde{n}) + 3p(\tilde{n})}{2\sqrt{13} \tilde{n}} \approx 203 \text{ MeV} \quad .
\end{equation}

The Oppenheimer-Volkoff equation describes the interior of a static star in general relativity:
\begin{equation} \label{Op-V} 
\frac{dp}{dr}=-\frac{\left( \rho + p \right)\left[m(r)+4\pi r^3 p \right]}{r\left[ r-2m(r) \right]} \quad ,
\end{equation}
where
\begin{equation} \label{stellarMass}
 m(r) = \int_0^r 4\pi \rho \left( \xi \right) \xi^2 d\xi \quad .
\end{equation}
The calculation of the maxmimum neutron star mass limit begins by choosing a central energy density value $\rho_c$ (or, equivalently, a central baryon density $n_c$) and integrating Eq. (\ref{Op-V}) radially outward to the edge of the star, marked by the radius $R$ where the pressure drops to zero. One of the maximally incompressible equations of state are used in the high density interior out to the radius where the energy density drops below the matching energy density $\tilde{\rho} = \rho\left( \tilde{n} \right)$. The experimentally verified low density equation of state is then used out to the edge of the star, and the mass of the star $M(R)$ determined from Eq. (\ref{stellarMass}). The integration of Eq. (\ref{Op-V}) is repeated for a series of progressively increasing central densities. The maximum mass value is determined by the central density for which $M\left( \rho_c \right)$ has its stationary point ($dM/d\rho_c = 0$). Above this central density $dM/d\rho_c < 0$ and the star will be unstable to collapse to a black hole from radial perturbations (see, for example, Refs. \cite{Shapiro83} or \cite{Glendenning96}).

Shown in Fig. 1 is the neutron star mass as a function of the central baryon density, with the BPS equation of state used below normal nuclear density, the WFF-A equation of state used above normal nuclear density and below the matching energy density, and the kinetic theory maximally incompressible equation of state Eq. (\ref{newMaxIncompressible}) used above the matching energy density. Mass models having a matching density corresponding to one, two, and four times normal nuclear energy density $\rho_{nn}$ are shown. Also shown for comparison are models with the sound speed equal to light speed maximally incompressible equation of state Eq. (\ref{oldMaxIncompressible}) of Refs. \cite{Rhoades74} and \cite{Kalogera96} used above the matching density, and models with the WFF-A equation of state used at all densities above normal nuclear density. Tables I-III list some specific numerical values of the maximum mass for a few different matching densities for the WFF-A, WFF-U, and APR equations of state. The softer kinetic theory maximally incompressible equation of state is seen to yield smaller maximum mass values than the sound speed equal to light speed maximally incompressible equation of state. If the WFF-A equation of state is regarded as experimentally verified up to twice normal nuclear energy density, as was assumed in Ref. \cite{Kalogera96}, the result of using the kinetic theory maximally incompressible equation of state instead of the sound speed equal to light speed maximally incompressible equation of state is to reduce the maximum neutron star mass from 2.92 $M_\odot$ to 2.63 $M_\odot$. The average maximum mass reduction for these three low density equations of state, when using a matching energy density of twice normal nuclear density, is 0.29 $M_\odot$, from an average of 2.93 $M_\odot$ to 2.64 $M_\odot$.

Figure 2 shows the maximum neutron star mass as a function of the matching energy density for the three low density equations of state and the two maximally incompressible high density equations of state. The three low density equations of state yield similar maximum mass values for matching densities up to twice normal nuclear energy density, so the value of the maximum mass derived from the kinetic theory maximally incompressible equation of state is substantially independent of the low density model for matching densities in this density range. Using the WFF-A, WFF-U, or APR equations of state for all densities above normal nuclear density yields a maximum mass of, respectively, 2.13, 2.19, or 2.21 $M_\odot$. Examination of Fig. 2 shows when the matching energy density approaches four times normal nuclear energy density, the stellar mass value is largely determined by the low density equation of state. As was noted by Kalogera and Baym \cite{Kalogera96}, once nuclear physics advances to where the experimentally based equation of state is verified up to four times normal nuclear density, the maximum neutron star mass problem will be solved.

\section{discussion}
The principal finding of this paper is stability and causality constraints resulting from applying kinetic theory to thermal and viscous dissipative processes in neutron star matter yields a softer maximally incompressible equation of state, and a lower maximum mass of neutron stars, than for the sound speed equal to light speed maximally incompressible equation of state. The reduction in the maximum mass averages from 2.93 $M_\odot$ down to 2.64 $M_\odot$ for the three low density equations of state considered if the low density equation of state is regarded as experimentally verified up to twice normal nuclear energy density.

The softening of the maximally incompressible equation of state is a generic prediction of kinetic theory, but the specific value of the resulting maximum neutron star mass derived here rests upon the accuracy of approximating the relaxation times and viscous-thermal couplings with their noninteracting degenerate Fermi gas forms, and the overall validity of dissipative relativistic kinetic theory for strongly interacting nuclear matter at high density. Future data from the Relativistic Heavy Ion Collider holds promise for experimental testing of the applicability of the kinetic theory approach to modeling high density nuclear matter.

Only static neutron star mass models have been constructed in this paper. Rotation can support larger mass stars against collapse, but only for stars rotating rapidly enough to be nearly shedding mass from the equator is the maximum mass value significantly changed from its nonrotating value. For these rapidly rotating stars the increase in the maximum mass is at most approximately 20\% over the static value when undergoing uniform rotation \cite{Cook92,Cook94}. Differential rotation is strongly damped \cite{Hegyi77}. The remnant of binary neutron star coalescence may undergo a short period of dynamically stable differential rotation, and thus briefly support a much larger mass neutron star remnant against prompt black hole collapse than is possible for a static or uniformly rotating star \cite{Baumgarter00}.

\acknowledgements
This work was supported by the U.S. National Aeronautics and Space Administration through Grant No. NAG5-9541.


\begin{table}
\caption{The maximum neutron star mass, in solar masses, with the WFF-A equation of state used at low density and a maximally incompressible equation of state used at high density.}
\begin{tabular}{ccccc}
&\multicolumn{4}{c}{Matching energy density}\\
&$\rho_{nn}$\tablenote{Normal nuclear energy density.}&$2\rho_{nn}$&$3\rho_{nn}$&$4\rho_{nn}$\\
\tableline
Luminal\tablenote{Sound speed equal to light speed maximally incompressible equation of state.}&4.08&2.92&2.45&2.22\\
Kinetic\tablenote{Kinetic theory maximally incompressible equation of state.}&3.68&2.63&2.22&2.03\\
\end{tabular}
\end{table}

\begin{table}
\caption{The maximum neutron star mass, in solar masses, with the WFF-U equation of state used at low density and a maximally incompressible equation of state used at high density.}
\begin{tabular}{ccccc}
&\multicolumn{4}{c}{Matching energy density}\\
&$\rho_{nn}$&$2\rho_{nn}$&$3\rho_{nn}$&$4\rho_{nn}$\\
\tableline
Luminal&4.08&2.93&2.48&2.27\\
Kinetic&3.68&2.65&2.25&2.10\\
\end{tabular}
\end{table}

\begin{table}
\caption{The maximum neutron star mass, in solar masses, with the APR equation of state used at low density and a maximally incompressible equation of state used at high density.}
\begin{tabular}{ccccc}
&\multicolumn{4}{c}{Matching energy density}\\
&$\rho_{nn}$&$2\rho_{nn}$&$3\rho_{nn}$&$4\rho_{nn}$\\
\tableline
Luminal&4.08&2.93&2.48&2.28\\
Kinetic&3.68&2.65&2.26&2.12\\
\end{tabular}
\end{table}

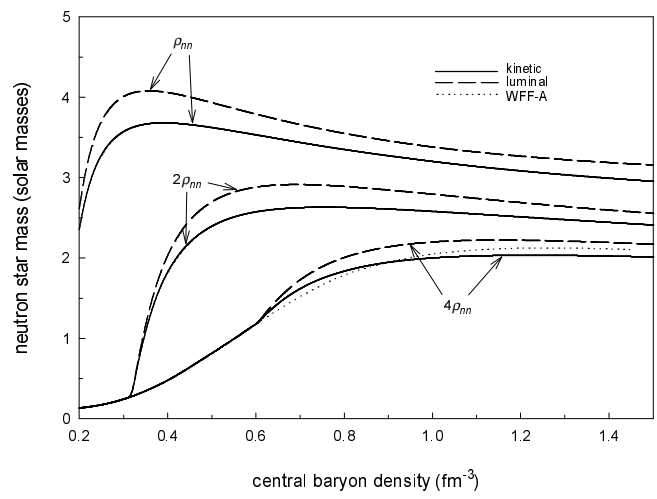
\begin{figure} \label{fig1}
\caption{The maximum neutron star mass as a function of the central baryon density and the matching energy density. The matching energy density is in units of the normal nuclear energy density $\rho_{nn}$. The lines labeled "kinetic" are for the WFF-A equation of state used at low density and the kinetic theory maximally incompressible equation of state used at high density. The lines labeled "luminal" are for the WFF-A equation of state used at low density and the sound speed equal to light speed maximally incompressible equation of state used at high density. The "WFF-A" line is for the WFF-A equation of state used for all densities above normal nuclear density.}
\end{figure}

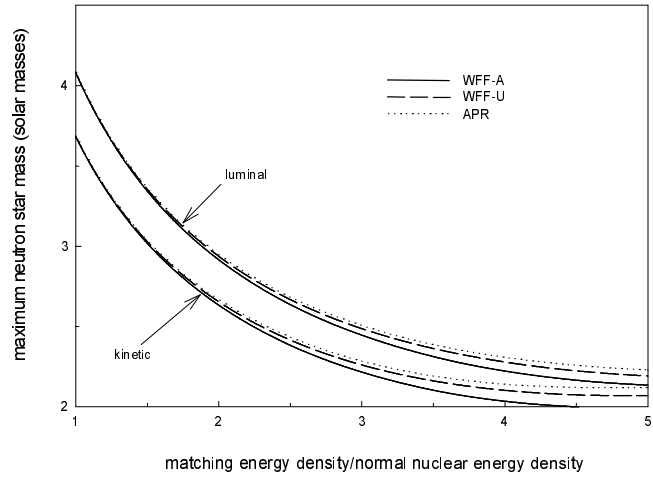
\begin{figure} \label{fig2}
\caption{The maximum neutron star mass as a function of the matching energy density for the WFF-A, WFF-U, or the APR equation of state used at low density. The lines labeled "kinetic" are for the kinetic theory maximally incompressible equation of state used at high density, and the lines labeled "luminal" are for the sound speed equal to light speed maximally incompressible equation of state used at high density.}
\end{figure}

\end{document}